\documentclass{article}
\usepackage{amsmath,amsthm}
\usepackage{graphicx,epstopdf,epsfig,multirow,epic,bm}

\normalsize \rm
\begin{document}

\begin{center}
{\Large \textbf { Random walks on Fibonacci treelike models: emergence of power law}}\\[12pt]
{\large Fei Ma$^{a,}$\footnote{~The author's E-mail: mafei123987@163.com. }, \quad  Ping Wang$^{b,c,d,}$\footnote{~The corresponding author's E-mail: pwang@pku.edu.cn.} \quad  and  \quad Bing Yao$^{e,}$\footnote{~The author's E-mail: yybb918@163.com.}}\\[6pt]
{\footnotesize $^{a}$ School of Electronics Engineering and Computer Science, Peking University, Beijing 100871, China\\
$^{b}$ School of Software and Microelectronics, Peking University, Beijing  102600, China\\
$^{c}$ National Engineering Research Center for Software Engineering, Peking University, Beijing, China\\
$^{d}$ Key Laboratory of High Confidence Software Technologies (PKU), Ministry of Education, Beijing, China\\
$^{e}$ College of Mathematics and Statistics, Northwest Normal University, Lanzhou  730070, China}\\[12pt]
\end{center}

\begin{quote}In this paper, we propose a class of growth models, named Fibonacci trees $F(t)$, with respect to the intrinsic advantage of Fibonacci sequence $\{F_{t}\}$. First, we turn out model $F(t)$ to have power-law degree distribution with exponent $\gamma$ greater than $3$. And then, we study analytically two significant indices correlated to random walks on networks, namely, both the optimal mean first-passage time ($OMFPT$) and the mean first-passage time ($MFPT$). We obtain a closed-form expression of $OMFPT$ using algorithm 1. Meanwhile, algorithm 2 and algorithm 3 are introduced, respectively, to capture a valid solution to $MFPT$. We demonstrate that our algorithms are able to be widely applied to many network models with self-similar structure to derive desired solution to $OMFPT$ or $MFPT$. Especially, we capture a nontrivial result that the $MFPT$ reported by algorithm 3 is no longer correlated linearly with the order of model $F(t)$.


\end{quote}

\vskip 1cm

\section{Introduction}

The problem of diffusion in disordered systems is of typical interest in various kinds of disciplines, including statistics physics, applied mathematics, theoretical computer science as well biological science and so forth \cite{Gomez-Gardenes-2018}-\cite{Holcman-2014}. Among such kind of diffusion, as a representation characterizing discrete-time diffusive processes, random walks have received considerable attention and still remain as an active research topic in the future \cite{Curado-2020}-\cite{Bollt-2005}.

More recently, complex networks have received tremendous concerns from different scientific communities \cite{Albert-1999-1}-\cite{M. E. J. Newman-2005}. In the last two decades, a lot of significant results related to complex networks have been obtained. Among them, particularly, two prominent characters of numerous complex networks have been unveiled, that is, the concept of small-world due to Watts and Strogatz \cite{Watts-1998} and the scale-free feature proposed by Barabasi and Albert \cite{Albert-1999-1}. The both above characters convince people that most complex networks are not completely disordered but seemingly ordered. Such intriguing rules behind complex networks are inspiring more and more researchers to devote themselves to deeply understanding structure and dynamics on networks \cite{J. Zhang-2006}-\cite{P. Hu-2018}.

The current issues in the study of complex networks mainly contain two aspects. The one is to generate available models to understand how complex networks evolve dynamically over time. The most significant of models of this kind is the BA-model \cite{Albert-1999-1} which is perfectly constructed based on two key mechanisms, \emph{growth} and \emph{preferential attachment}. After that, a large variety of network models, including stochastic and deterministic, have been proposed to intend to stimulate evolutional principle on real-world complex networks \cite{Dorogovtsev-2002,Dorogovtsev-2000}. The other is to focus on learning how some intriguing dynamical behaviors are influenced by the underlying topological structure of complex networks \cite{Dorogovtsev-2008,Liu-2016}. Discussed behaviors occurring on complex networks have: (1) synchronization phenomena \cite{Turalska-2010}, (2) chaotic control for system identification \cite{Carroll-2004}, (3) percolation \cite{ Li-2013}, (4) epidemic spreading \cite{Valdez-2018} and (5) random walks \cite{Wijesundera-2016} and so on. In order to better understand the aforementioned behaviors, various type of models need to be generated. Several well-known deterministic models of them are Apollonian networks \cite{Andrade-2005}, Sierpinski network \cite{Wang-2017}, Recursive tree network \cite{Kursten-2016}, Cayley tree network \cite{Li-2015} and Regular lattice model \cite{Hein-2010}. Contrasted with the first three models, the later two have no the scale-free feature.

Motivated by the descriptions above, we here propose a class of new deterministic models, named Fibonacci trees $F(t)$. Our aims are not only to report whether models $F(t)$ have power-law distribution but also to study two types of random walks on models $F(t)$ in detail. Compared with numerous published network models, our models $F(t)$ are built based on the famous Fibonacci sequence $\{F_{t}\}$. This is reason that we call models $F(t)$ Fibonacci trees. Appendix A gives a brief introduction to Fibonacci sequence $\{F_{t}\}$. In addition, different from both Apollonian networks and Sierpinski network, models $F(t)$ have a tree structure and are available to research some physical behaviors of great interest. At the same time, models $F(t)$ have the scale-free feature popular in a great deal of complex networks in comparison with both Cayley tree network \cite{Li-2015} and Regular lattice model. This implies that our models are heterogeneous and the later both have homogeneous structure. Although the scale-free feature can be found on both Recursive tree network and models $F(t)$ simultaneously, the construction of models $F(t)$ differs in that of the former mainly because Fibonacci sequence $\{F_{t}\}$ has some advantage, for instance, seeing our recent works associated with Fibonacci sequence $\{F_{t}\}$ for more details \cite{F.M-2018-1,F.M-2018-2}.

The rest of this paper can be organized by the following several sections. First of all, we take an introduction to Fibonacci trees $F(t)$ in Section 2. Beside that, the basic structural parameters, i.e., vertex number and edge number, can be analytically obtained. Based on the above parameters and specific structure of Fibonacci trees $F(t)$, we in this section state that models $F(t)$ have power-law distribution with exponent $\gamma$ no less than $3$. Next, Section 3 mostly contains our main discussions about two different kinds of random walks on models $F(t)$. In order to better understand random walks, some related work has to be recalled. And then, we provide algorithm 1 for computing a concise solution to the optimal mean first-passage time for random walks on models $F(t)$ according to distance between trapping vertex and source vertex. Furthermore, to calculate solution of the mean first-passage time for random walks on models $F(t)$, two effective algorithms are adopted. The one is for deriving an approximate solution and the other for an exact solution. After obtaining desirable results, we give a plausible and heuristic explanation between efficiency of diffusion about random walks and the underlying topological structure of models $F(t)$. For the concrete outline of this paper, we have to close this paper by making an elaborated conclusion and bringing some meaningful problems in the last section.

\section{Description of Fibonacci trees $F(t)$}

\subsection{Construction}

\begin{figure}
\centering
  \includegraphics[height=8cm]{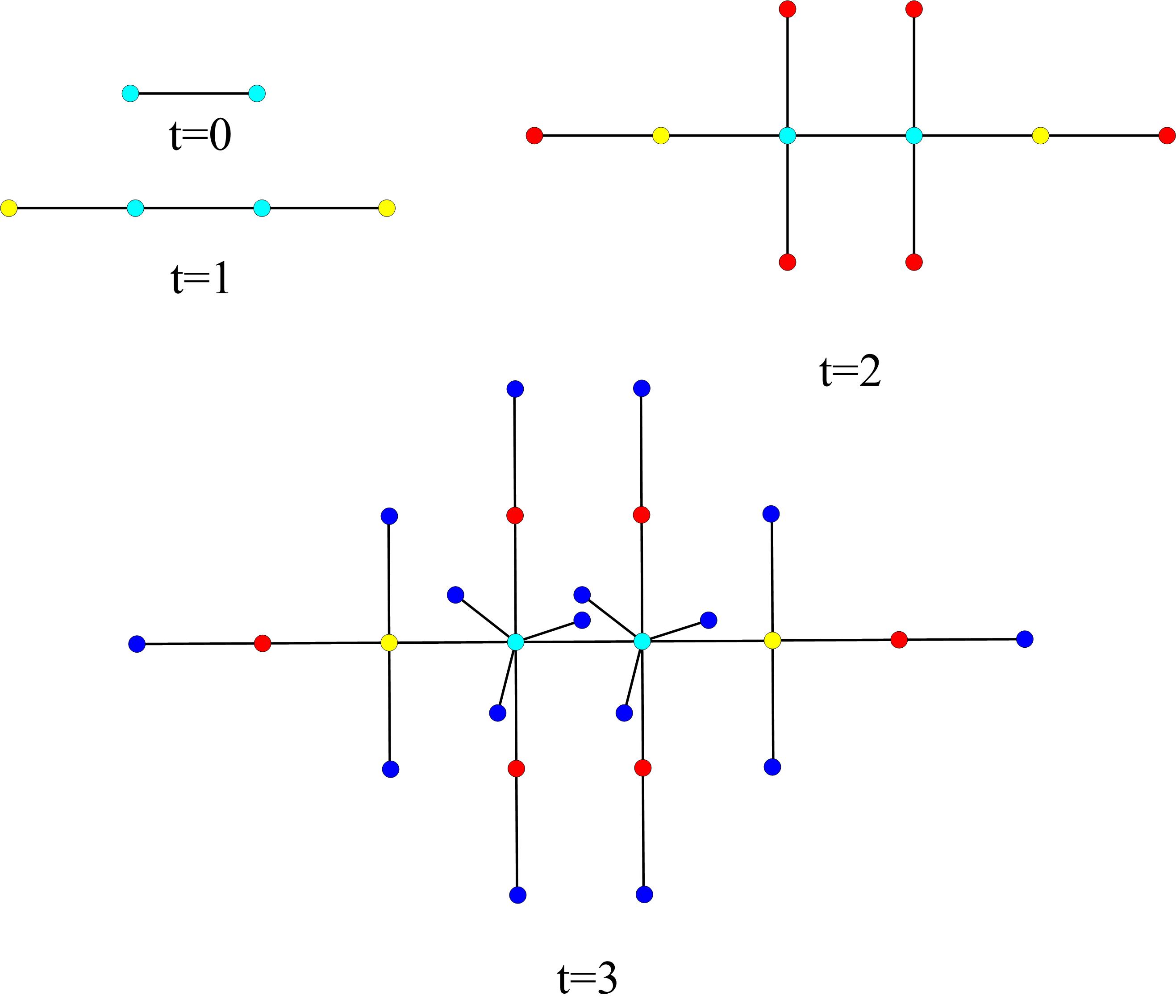}\\
{\small Fig.1. The diagram of first several generations of Fibonacci trees $F(t)$.  }
\end{figure}

We, in this section, will introduce a family of Fibonacci trees $F(t)$ which can be used to study many intriguing topics on complex network models, such as scale-free feature, small-world property, and the mean first-passage time ($MFPT$) for random walks, as well as the synchronization phenomena. First of all, the generation of growth Fibonacci trees may be easy to understand using the below algorithm, as follows

As $t=0$, the seminal model, called $F(0)$, is an edge connecting two initial vertices of degree one. For convenience, let the degree of each vertex of $F(0)$ be $F_{0}=1$ similar to the first term of Fibonacci sequence $\{F_{t}\}$, see Appendix A.

As $t=1$, we add two new vertices which are separately linked with each vertex of $F(0)$, resulting in the next model $F(1)$. Put another way, the model $F(1)$ can be obtained from the $F(0)$ by connecting $F_{1}$ new vertex to each degree $F_{0}$ vertex. Such a description is the same as the property of Fibonacci sequence $\{F_{t}\}$ and then is adopted to produce the future models.

As $t\geq 2$, the model $F(t)$ can be obtained from the $F(t-1)$ by creating $F_{t-i}$ new vertices for each vertex $v$ added at time step $i$ into the $F(t-1)$ and connecting the $F_{t-i}$ new vertices to vertex $v$ for $i=0, 1,2,...,t-1$.

Figure 1 depicts the first several generations of the desired Fibonacci trees $F(t)$. After that, based on the above description of Fibonacci trees $F(t)$, both the order $|V(t)|$ (number of vertices) and the size $|E(t)|$ (number of edges) of Fibonacci trees $F(t)$ can be easily computed. Before computing, taking into consideration the relationship between our model $F(t)$ and Fibonacci sequence $\{F_{t}\}$, we need to utilize some helpful notations. The $i-root$ subtree $ST_{i}(t)$ of model $F(t)$ is a subtree of Fibonacci trees $F(t)$ which contains all vertices generated on the basis of root vertex $i$. As an illustrative example, several smaller $i-root$ subtrees $ST_{i}(3)$ is shown in Fig.2. The symbol $|V(t,i)|$ represents the number of vertices of $i-root$ subtree $ST_{i}(t)$. Obviously, $|V(t)|=2|V(t,0)|$. If let $\mathbf{V(t)^{i}_{1}}=(|V(t,1)|,|V(t,2)|,...,|V(t,i)|)^{T}$ be an $i$-dimensional vector and $\mathbf{1(i)}=(1,1,...,1)^{T}$ an $i$-dimensional identity vector where the superscript $T$ indicates permutation, then we can write the vector representation among $|V(t,i)|$ of Fibonacci tree $F(t)$

\begin{equation}\label{eqa:2-1-1}
\mathbf{V(t)^{t-1}_{0}}-\mathbf{1(t)}=\mathbf{F^{\Delta}(t)}\mathbf{V(t)^{t}_{1}}
\end{equation}
here $\mathbf{F^{\Delta}(t)}=(f_{ij})$ is a $t$-dimensional square matrix in which $f_{ij}=F_{j-i+1}\;(i\leq j\leq t)$, shown in Appendix B. With several initial conditions of $|V(t,i)|$, namely $|V(t,t)|=1, |V(t,t-1)|=2$, and $|V(t,t-2)|=5$, one can obtain the solution of $|V(t,i)|$ ($0\leq i\leq t-2$)

\begin{equation}\label{eqa:2-1-2}
|V(t,i)|=\left\{\begin{split}
&\frac{1+(7+4\sqrt{3})(\sqrt{3}+1)^{t-i-2}+(7-4\sqrt{3})(1-\sqrt{3})^{t-i-2}}{3}, \qquad 0\leq i\leq t-2 \\
&\qquad2, \qquad\qquad i=t-1 \\
&\qquad1, \qquad\qquad i=t \\
\end{split}
\right.
\end{equation}

Therefore, after the time step $t$, the order $|V(t)|$ of Fibonacci tree $F(t)$ is equal to $$2[1+(7+4\sqrt{3})(\sqrt{3}+1)^{t-2}+(7-4\sqrt{3})(1-\sqrt{3})^{t-2}]/3.$$

\subsection{Scale-free feature}

After Barab\'{a}si and Albert \cite{Albert-1999-1} first proposed the scale-free feature popular in man-made and natural complex networks, a large number of related works have been done. One main topic of these researches is to construct available models (stochastic and deterministic) to explain the reasons for emergence of the scale-free feature. For any given deterministic model, whether it is scale-free or not can be easily verified using vertex-degree distribution. We here provide a brief explanation to the power-law property of Fibonacci trees $F(t)$. Before proving, it is helpful to bring some notations. Let $\Delta V(i)$ denote by the set of vertices created at time step $i$, the corresponding symbol $|\Delta V(i)|$ is the number of vertices belonging to the set $\Delta V(i)$. Most obviously, $|\Delta V(0)|=2$ and $|\Delta V(1)|=2$ can be obtained by hand. In addition, $\mathbf{V^{i}_{1}}=(|\Delta V(1)|,|\Delta V(2)|,...,|\Delta V(i)|)^{T}$ is defined as a vector with dimension $i$ where the superscript $T$ represents permutation as before. Based on the description above, we can have

\begin{figure}
\centering
  \includegraphics[height=5.5cm]{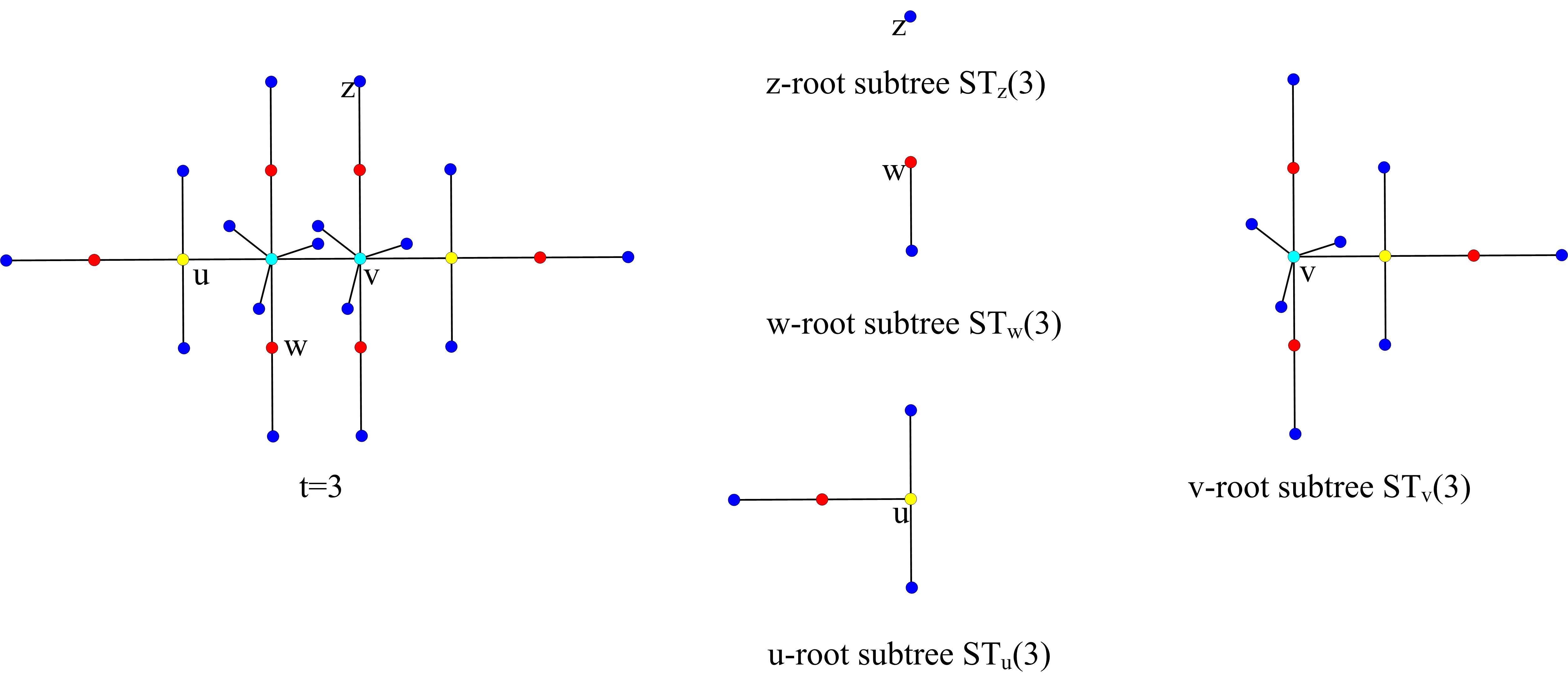}\\
{\small Fig.2. The diagram of several smaller $i-root$ subtrees $ST_{i}(3)$ on Fibonacci tree $F(3)$. We label each kind of color vertices of model $F(3)$ using distinct letters, such as yellow vertex with letter $u$, indigo vertex with letter $v$, red vertex with letter $w$ and blue vertex with letter $z$. Thus, Fibonacci tree $F(3)$ has two $v-root$ subtrees $ST_{v}(3)$, two $u-root$ subtrees $ST_{u}(3)$, six $w-root$ subtrees $ST_{w}(3)$ as well as sixteen $z-root$ subtrees $ST_{z}(3)$. Note that, the $z-root$ subtree $ST_{z}(3)$ is an empty tree which contains one isolated vertex.   }
\end{figure}

\begin{equation}\label{eqa:2-2-1}
\mathbf{V^{t}_{1}}=\mathbf{F^{\Delta}(t)}\mathbf{V^{t-1}_{0}}
\end{equation}
in which $\mathbf{F^{\Delta}(t)}$ has a similar meaning as mentioned in Eq.(\ref{eqa:2-1-1}), see Appendix B for more details. After implementing some simple matrix arithmetics, the values of $|\Delta V(i)|$ satisfy the below equations

\begin{equation}\label{eqa:2-2-2}
|\Delta V(t)|=\left\{\begin{split}&\qquad 2,\qquad\qquad  t=0 \\
&\qquad2, \qquad\qquad t=1 \\
&\frac{(9+5\sqrt{3})(\sqrt{3}+1)^{t-2}+(9-5\sqrt{3})(1-\sqrt{3})^{t-2}}{3}, \qquad t\geq2 \\
\end{split}
\right.
\end{equation}

Next, through the growth process of our models $F(t)$, the degree $k_{i}(t)$ of each vertex added at time step $i$ can be written in the following form $$k_{i}(t)=\sum_{j=0}^{t-i}F_{j}=\frac{1}{\sqrt{5}}\left[\left(\frac{\sqrt{5}+1}{2}\right)^{t+2-i}-\left(\frac{\sqrt{5}-1}{2}\right)^{t+2-i}\right].$$

With the help of relations $|\Delta V(i)|$ to $k_{i}(t)$, the degree distribution of Fibonacci trees $F(t)$ can be expressed as follows

\begin{equation}\label{eqa:2-2-3}
P_{cum}(k\geq k_{i})=\frac{\sum_{j\leq i}|\Delta V(j)|}{|V(t)|}\propto k_{i}^{-\frac{\ln(1+\sqrt{3})}{\ln(1+\sqrt{5})-\ln2}}.
\end{equation}

Interestingly, the degree exponent $\gamma$ of model $F(t)$ does not fall into the typical scope between $2$ and $3$ but is larger than $3$, namely,
$$\gamma=1+\frac{\ln(1+\sqrt{3})}{\ln(1+\sqrt{5})-\ln2}> 3.$$

To the best of our knowledge, almost all the published deterministic models have degree exponents $\gamma$ no more than $3$, such as Apollonian network with power-law exponent equal to $1+\frac{\ln 3}{\ln 2}<3$ \cite{Zhang-2009}, Sierpinski networks with power-law exponent equivalent to $2+\frac{\ln 2}{\ln 3}<3$ \cite{Zhang-2007}. In contrast with such scale-free models, our model $F(t)$ has its own advantages in the generation process. We in this paper take a new concept of preferential attachment, called velocity-based growth, for the construction of Fibonacci trees $F(t)$. Such a manner is completely different from that of commonly utilized in \cite{Zhang-2009}-\cite{ F-J-Y-2018}. The later models are built based absolutely on vertex degree, that is to say, the changed quantity $\delta_{v}(t)$ of the degree of vertex $v$ at two consecutive phases following $\delta_{v}(t)=k_{v}(t)-k_{v}(t-1)$ $(t\geq1)$, and used to simulate discrete Markov property. However the mechanism of our models $F(t)$ can be referred to as another statement, namely $\delta_{v}(t)=F_{t}$. In other words, the widely discussed models has growth rate $\delta_{v}(t)/k_{v}(t-1)=\alpha$ where $\alpha$ is an invariable integer. The growth rate of Fibonacci trees $F(t)$ is not a constant but a variable. For the large value of $t$, $\delta_{v}(t)/k_{v}(t-1)= F_{t}/F_{t+1}\approx 0.618$. As a result, model $F(t)$ can be conventionally called \emph{golden section tree model}. Maybe the reason for degree exponent $\gamma$ of Fibonacci trees $F(t)$ no less than 3 is the fresh growth manner, which will be a topic to deeply study in our future research.

\section{Random walks on Fibonacci trees $F(t)$}

As known, information diffusion over complex networks is one of the most significant research interests among different kinds of disciplines, including statistic physics, theoretical computer science, applied mathematics, and chemistry as well as biology \cite{Nematzadeh-2014}-\cite{Iribarren-2009}. The widely used statistical parameters for depicting the efficiency of information diffusion mainly consists of diameter, average path length, unbiased Markov random walk, and biased Markov random walk. In this section, we will pay more attention to study the later two indices and make a detailed comparison between them. In order to organize the rest of this section smoothly, we should give some helpful concepts and terminologies used later.

Given a complex structural model $\mathbf{M}$, at each time, a designated walker (particle) over model $\mathbf{M}$ is able to move from its present location to arbitrary of its nearest neighbors with equal probability. Such a motion is defined to as the unbiased Markov random walk which is usually discussed using discrete-time Markov chain. Furthermore, if there is a destination object $o$ allocated at some vertex of model $\mathbf{M}$ which attracts all walkers to visit it, the problem addressed above becomes the typical trapping problem with a unique absorber \cite{Montroll-1969}. A key index characterizing this type of trapping problem is mean first-passage time, shorted $MFPT$, in terms of $MFPT=\frac{1}{|\mathbf{M}|-1}\sum_{j\in V, j\neq o}L_{j}$ where $L_{j}$ is the first-passage time ($FPT$) for a walker starting from a position $j$ and $|\mathbf{M}|$ is the vertex number as well as $o$ represents this trap. Broadly speaking, the solution to $MFPT$ can be attained using the fundamental matrix method \cite{Kemeny-1976}. Here, for the above model $\mathbf{M}$, its adjacency matrix $\mathbf{A}=(a_{ij})$ is as follows: $a_{ij}$ is equal to $1$ if the pair of nodes $i$ and $j$ is connected by an edge, otherwise $a_{ij}$ is zero. By definition, the degree $k_{i}$ of vertex $i$ equals $\sum_{j=1}^{|\mathbf{M}|}a_{ij}$. And then, the concrete diagonal degree matrix $\mathbf{Z}$ is written as $\mathbf{Z}=diag(k_{1}, k_{2}, ..., k_{|\mathbf{M}|})$. According to these descriptions, the normalized Laplacian matrix of $\mathbf{M}$ is $\mathbf{L}=\mathbf{I}-\mathbf{Z}^{-1}\mathbf{A}$ whose entity $l_{ij}=1-a_{ij}/k_{i}$ and where symbol $\mathbf{I}$ is an identity matrix. Based on the basic materials, the transition probability for a walker seated at vertex $i$ of model $\mathbf{M}$ to jump to any vertex $j$ by exactly one step is clearly denoted by $a_{ij}/k_{i}$. Therefore, for a walker starting from vertex $i$ at $t=0$, the probability $P_{ij}(t)$ that it spends $t$ time steps removing from $i$ to $j$ ought to satisfy the below master equation

\begin{equation}\label{eqa:3-0-1}
P_{ij}(t)=\sum_{w=1}^{|\mathbf{M}|}\frac{a_{wj}}{k_{w}}P_{iw}(t-1).
\end{equation}

Similarly, for this walker, time $L_{ij}(t)$ taken to first jump to vertex $j$ is

\begin{equation}\label{eqa:3-0-2}
L_{ij}(t)=\sum_{w\in \mathbf{M}, w\neq j}\frac{a_{wj}}{k_{w}}L_{iw}(t-1)+1.
\end{equation}

Since then, the solution to $MFPT$ is reduced to compute the following equation

\begin{equation}\label{eqa:3-0-3}
MFPT=\frac{1}{|\mathbf{M}|-1}\sum_{j\in \mathbf{M}, j\neq o}L_{jo}=\frac{1}{|\mathbf{M}|-1}\sum_{j\in \mathbf{M}, j\neq o}\sum_{v\in \mathbf{M}, v\neq o}\left(\frac{a_{vo}}{k_{v}}L_{jv}+1\right).
\end{equation}

As said in reference \cite{Zhang-2009}, Eq.(\ref{eqa:3-0-3}) can be further expressed as

\begin{equation}\label{eqa:3-0-4}
MFPT=\frac{1}{|\mathbf{M}|-1}\sum_{j\in \mathbf{M}, j\neq o}\sum_{v\in \mathbf{M}, v\neq o}(\Delta^{-1})_{jv}.
\end{equation}
here $(\Delta^{-1})_{jv}$ is an entity of a resulting matrix of the normalized discrete Laplacian matrix $\mathbf{L}$ in which both row and column corresponding to the trap vertex $o$ have been deleted. Interested reader is referred to \cite{Zhang-2009} for more detail. From the brief description above, it is not difficult to obtain the closed-form expression to $MFPT$ for random walks on many models with a few vertices. However, for lots of models with thousands of vertices, such work seems to be tough and impracticable because of tremendous demands to both time consumption and space memory. Even though there are challenges for concerning random walks on networks, its importance can not be ignored and hence some well-known deterministic models have been proposed to probe many interesting properties related to random walks. Included models have Vicsek fractals \cite{Zhang-2010}, T-fractal \cite{Agliari-2008} and deterministic recursive trees \cite{Comellas-2010}.

The unbiased Markov random walk emphasizes a type of ideal situation in which a walker has no information of the model $\mathbf{M}$ and just chooses uniformly at random one vertex of its neighbor set as a candidate. It is in fact possible for a walker to attempt to spend as less time as possible to arrive at that designed trap. In this case, the distance between that trap and source vertex at which a walker is initially located will play a vital role during the transition process. The distance $d_{ij}$ of a pair of vertices, say $i$ and $j$, is the number of edges belonging to one of the shortest paths between vertex $i$ and vertex $j$. Compared with the universal situation with equal probability, another environment is that a walker has all information over the entire model $\mathbf{M}$, i.e., knowing the shortest path from its current position to that specific trap. Such a transition process can be called the biased Markov random walk or the optimal Markov random walk. For a model $\mathbf{M}$ under consideration, the first-passage time for a walker who completely understands to take some shortest path to visit its destination $o$ may be defined to be the optimal $FPT$, sometimes abbreviated to $OFPT$. For the whole model $\mathbf{M}$, the term $OMFPT$ obeys

\begin{equation}\label{eqa:3-0-5}
OMFPT=\frac{1}{|\mathbf{M}|-1}\sum_{j\in \mathbf{M}, j\neq o}d_{jo}.
\end{equation}

From now on, let us put insights into discussing both the unbiased Markov random walk and the optimal Markov random walk on our models $F(t)$. Below is our main works consisting of providing algorithm 1 for $OMFPT$, algorithm 2 for $MFPT$, and a closed-form solution to $MFPT$ using algorithm 3 based on a series of recurrence equations.

\subsection{Algorithm 1 for $OMFPT$}

As a commonly studied graph model, tree itself has some advantages in comparison with other graph models. Here, one of them may be utilized to help us accomplish the computation of $OMFPT$, namely only a unique path connecting a pair of different vertices on tree. Before proceeding further, we should reconstruct Fibonacci trees $F(t)$ in a novel manner which is certainly on the basis of the intrinsic character of Fibonacci sequence $\{F_{t}\}$. As illustrated in Fig.3, models $F(t)$ can be reorganized from four small components, namely two $F(t-1)$s and two $F(t-2)$s, in a way described as below. In order to smoothly state, we denote two largest degree vertices of $F(t)$ by $H_{1}(t)$ and $H_{2}(t)$. According to such a description, we first may be able to emerge one of two vertices $H_{i}(t-1)$ ($i=1,2$) on $F(t-1)$ and any vertex $H_{i}(t-2)$ ($i=1,2$) on $F(t-2)$ into a new vertex $\Theta_{1}(t)$ which will be one hub vertex of the next model $F(t)$. Apply similar procedure to the left both models $F(t-1)$ and $F(t-2)$ to create the other vertex $\Theta_{2}(t)$. Finally, we can take a new edge to link new vertex $\Theta_{1}(t)$ with vertex $\Theta_{2}(t)$ to produce model $F(t)$. After accomplishing the procedure, both alternative vertices $\Theta_{i}(t)$ ($i=1,2$) will formally become hub vertices $H_{i}(t)$ ($i=1,2$) on the youth model $F(t)$.

\begin{figure}
\centering
  \includegraphics[height=3.5cm]{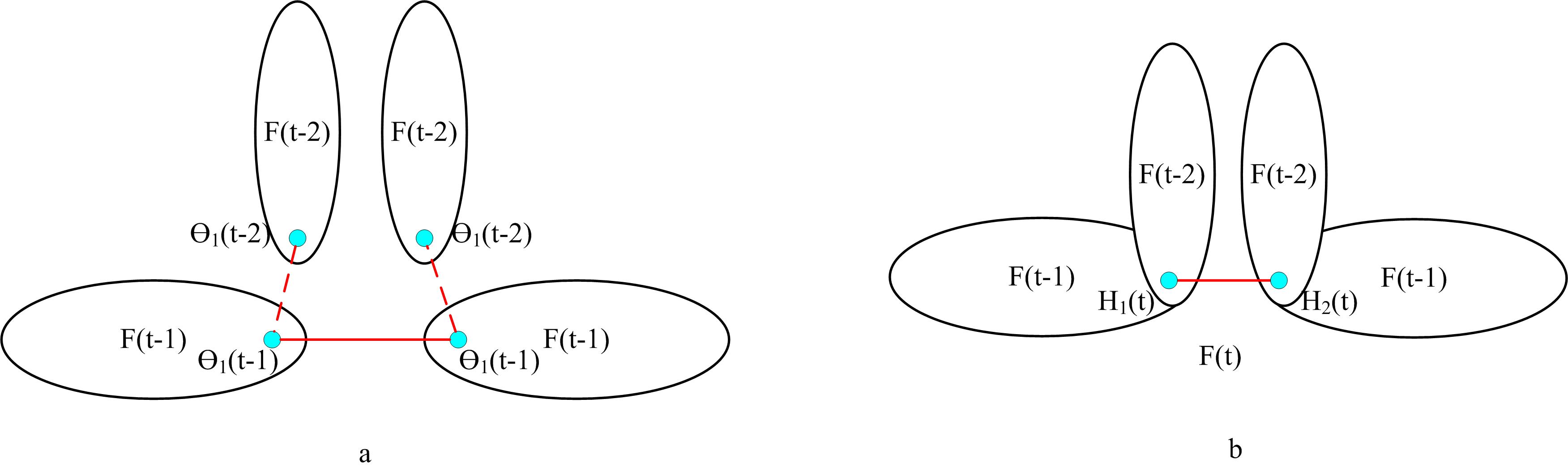}\\
{\small Fig.3. The diagram of another construction of Fibonacci tree $F(t)$.  }
\end{figure}

Given an attractor $o$ put at another of two hub vertices $H_{i}(t)$ ($i=1,2$) on model $F(t)$, our next aim is to calculate the concise solution to $OMFPT$. Taking into consideration the symmetric structure of model $F(t)$, we suppose that this attractor $o$ is seated at hub vertex $H_{1}(t)$ and then take use of symbol $OL(t)$ to indicate the total amount of time spent by arbitrary vertex to visit attractor $o$ along its own shortest path on model $F(t)$. Originally, it is not hard for one to obtain $OL(0)=1, OL(1)=4$, and $OL(2)=15$ by hand.

Thanks to the symmetric structure of model $F(t)$, at time step $t$, we have

\begin{equation}\label{eqa:3-1-1}
OL(t)=2(OL(t-1)+OL(t-2))+|V(t,0)|.
\end{equation}

Taking useful advantage of approach to non-homogeneous recurrence equations writes

\begin{equation}\label{eqa:3-1-2}
OL(t)=a_{t-1}OL(1)+b_{t}OL(0)+\sum_{i=0}^{t-2}a_{i}\Delta(t-2-i)
\end{equation}
in which $a_{t}=\frac{\sqrt{3}(1+\sqrt{3})^{t+1}-\sqrt{3}(1-\sqrt{3})^{t+1}}{6}$,  $b_{t}=\frac{\sqrt{3}(1+\sqrt{3})^{t-1}-\sqrt{3}(1-\sqrt{3})^{t-1}}{3}$ and $\Delta(t-2-i)=|V(t-i,0)|$.

Considering the initial conditions $OL(0)=1$ and $OL(1)=4$, Eq.(\ref{eqa:3-0-5}) is
resolved to yield
\begin{equation}\label{eqa:3-1-3}
OMFPT=\frac{1}{|V(t)|-1}\sum_{j\in V(t), j\neq o}d_{jo}=\frac{OL(t)}{|V(t)|-1}= O(t).
\end{equation}

Thus, we can obtain the following theorem

\textbf{Theorem 1} The solution of the optimal mean first-passage time ($OMFPT$) for trapping point located at either of two hubs of Fibonacci tree $F(t)$ obeys
\begin{equation}\label{eqa:3-1-a}
OMFPT=O(t).
\end{equation}

It is easy to see that the diameter $D(t)$ of our model $F(t)$ is equivalent to $2t+1$. Combining the result of Eq.(\ref{eqa:3-1-3}) and the expression of diameter of our model $F(t)$ implies a prominent fact that the diameter will play a crucial role on information transition process on model $F(t)$ when knowing information about all the shortest paths. As universally understood, almost all complex networks in nature and society have small-world property, i.e., relatively smaller diameter compared with the vertex number. Therefore, it is clear that a piece of information may be rapidly diffused over the whole network using less time. On the one hand, good news should first be fast spread to the hub vertex of network and in turn the hub vertex will forecast this news over the whole network in a similar manner. On the other hand, people should take effective measures to prevent network from attacks from bad news, such as rumor, virus. Such research is significantly dependent of network itself topological structure and has attracted more attention from many science communities. Here just introduces one of prominent indices figuring out information diffusion over complex networks under capturing knowledge about all the shortest paths. More usually, the situation mentioned above is rare. Since then, we will in the next subsection study random walks under other environment where a walker must choose uniformly at random one vertex of its neighbor set as its next hopping position to successfully arrive at its destination ultimately.

In order to discuss $MFPT$ for random walks on models $F(t)$ in detail, the below both subsections will mainly provide two algorithms, one is approximate and the other deterministic. We think that they may be conveniently used to obtain corresponding solution to $MFPT$ for random walks on other deterministic models with similar properties to models $F(t)$.

\begin{figure}
\centering
  \includegraphics[height=9cm]{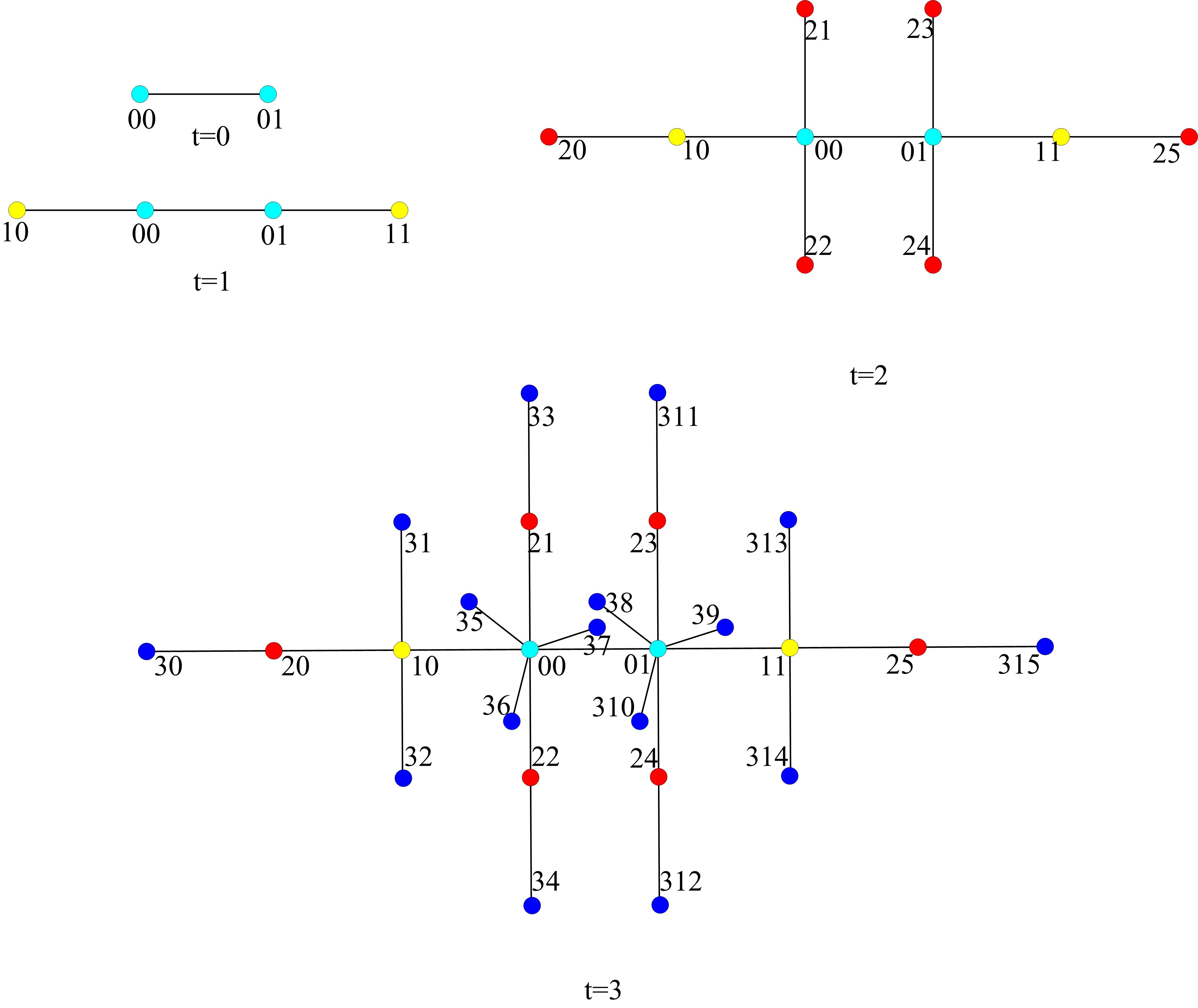}\\
{\small Fig.4. The diagram of reasonable labels on the first several Fibonacci trees $F(t)$.  }
\end{figure}

\subsection{Algorithm 2 for $MFPT$ }

To establish algorithm 2, we first give a unique label (like $ta$) to each vertex with respect to its generation. The first letter $t$ represents the time step when a vertex is created. The other $a$ is an ID number used to distinguish all vertices generated at the same time step. The first several generations have be reasonably labeled as outlined in Fig.4. The success of our Algorithm 2 for $MFPT$ indeed depends on such a labeling program. Based on Eq.(\ref{eqa:3-0-2}), below is a helpful table listing $FPT$ for each vertex on models $F(0)$, $F(1)$, $F(2)$ and $F(3)$, respectively. Here symbol $T_{ti}(t)$ indicates the $FPT$ for vertex labeled $ti$ on model $F(t)$.

\small \textbf{Table 1.}
\begin{center}
    \begin{tabular}{c c c c c c   }
       \hline
       $T_{00}(0)=0$ & $T_{00}(1)=0$ &  $T_{00}(2)=0$ &  $T_{00}(3)=0$  & $T_{01}(0)=1$  & $T_{01}(1)=3$  \\
      \hline
       $T_{01}(2)=9$ & $T_{01}(3)=25$ & $T_{10}(1)=1$ & $T_{10}(2)=3$ &  $T_{10}(3)=9$ &  $T_{11}(1)=4$  \\
       \hline
       $T_{11}(2)=12$ &  $T_{11}(3)=34$ & $T_{20}(2)=4$   & $T_{20}(3)=12$ & $T_{21}(2)=1$ & $T_{21}(3)=3$  \\
       \hline
       $T_{22}(2)=1$ &  $T_{22}(3)=3$ & $T_{23}(2)=1$ &  $T_{23}(3)=3$ & $T_{24}(2)=1$   & $T_{24}(3)=3$  \\
       \hline
       $T_{25}(2)=13$ & $T_{25}(3)=37$ &  $T_{30}(3)=13$ &  $T_{31}(3)=10$ & $T_{32}(3)=10$ &  $T_{33}(3)=4$  \\
       \hline
       $T_{34}(3)=4$   & $T_{35}(3)=1$ & $T_{36}(2)=1$ & $T_{37}(3)=1$ &  $T_{38}(3)=1$ &  $T_{39}(3)=1$ \\
       \hline
       $T_{310}(3)=1$ &  $T_{311}(3)=4$ & $T_{312}(3)=4$   & $T_{313}(3)=35$ & $T_{314}(2)=35$ & $T_{315}(3)=38$ \\
       \hline
     \end{tabular}
\end{center}

Seemingly, there is a rule behind the above table. To recover it, we must make use of some flexible techniques. In model $F(t)$, there are $F_{t-i}$ new vertices on degree one connected with a degree $k_{i}(t)=\sum_{j=0}^{t-i}F_{j}$ vertex $v$. For such a vertex $v$, we may define the transmit time from it to arbitrary one of its $F_{t-i}$ new vertices to as $\alpha$, and let $\beta$ denote the transmit time from vertex $v$ to arbitrary one of its $\sum_{j=0}^{t-1-i}F_{j}$ old vertices. With the help of Eq. (\ref{eqa:3-0-2}), we can have a couple of equations

\begin{equation}\label{eqa:3-2-1}
\alpha=1+\beta, \quad\beta=\frac{\sum_{j=0}^{t-1-i}F_{j}}{\sum_{j=0}^{t-i}F_{j}}+\frac{F_{t-i}}{\sum_{j=0}^{t-i}F_{j}}(1+\alpha).
\end{equation}

Thanks to Fibonacci sequence, the solution to $\beta$ can be easily obtained as below

\begin{equation}\label{eqa:3-2-2}
\beta=\frac{2F_{t-i}+\sum_{j=0}^{t-1-i}F_{j}}{\sum_{j=0}^{t-i}F_{j}}.
\end{equation}

With the large value for $t$, $\beta$ will coverage to a constant, i.e., $\beta\approx \sqrt{5}$. It appears to be proven in table 1. Because the convergence of the element $F_{i}$ of Fibonacci sequence $\{F_{t}\}$ is not monotone, $\beta$ also has an asymptotical trend to $\sqrt{5}$. Eq.(\ref{eqa:3-2-2}) in practice highlights a fact that the transmit time $T_{t_{i}j}(t)$ for a vertex $v$ obeys $T_{t_{i}j}(t)\approx\beta T_{t_{i}j}(t-1)$ as $t\rightarrow \infty$. On the basis of such an useful result, the algorithm 2 for $MFPT$ can be developed. Before doing this, we here employ symbol $T(i,t)$ to represent the transmit time sum of the total of vertices added at time step $i$ over model $F(t)$. Therefore, one may write

\begin{equation}\label{eqa:3-2-3}
T(t,t)=\sum_{i=0}^{|\Delta V(t)|-1}T_{ti}(t).
\end{equation}

Because of model $F(t)$ with a tree structure, plugging $\beta\approx \sqrt{5}$ into Eq.(\ref{eqa:3-2-3}) yields

\begin{equation}\label{eqa:3-2-4}
T(t,t)=|\Delta V(t)|+\sum_{i=0}^{t-1}F_{t-i}T(i,t)\approx|\Delta V(t)|+
\sum_{i=0}^{t-1}(\sqrt{5})^{t-i}F_{t-i}T(i,i)
\end{equation}
here $|\Delta V(t)|$ is the number of vertices added at time step $t$ into model $F(t)$, as described in Subsection 3.2. By induction, one may obtain

\begin{equation}\label{eqa:3-2-5}
T(t,t)\approx\frac{5+3\sqrt{5}}{2}T(t-1,t-1)+|\Delta V(t)|-\frac{5+\sqrt{5}}{2}|\Delta V(t-1)|.
\end{equation}

Apparently, applying some necessary methods for non-homogeneous recurrence equations to Eq.(\ref{eqa:3-2-5}) produces

\begin{equation}\label{eqa:3-2-6}
T(t,t)\approx\left(\frac{5+3\sqrt{5}}{2}\right)^{t}T(0,0)+\sum_{i=0}^{t-1}\left(\frac{5+3\sqrt{5}}{2}\right)^{i}\left[|\Delta V(t-i)|-\frac{5+\sqrt{5}}{2}|\Delta V(t-1-i)|\right].
\end{equation}

Based on Eqs.(\ref{eqa:3-0-3}) and (\ref{eqa:3-2-6}), an approximate solution to $MFPT$ can be expressed as

\begin{equation}\label{eqa:3-2-7}
MFPT_{t}=\frac{1}{|V(t)|-1}\sum_{j\in V(t), j\neq o}L_{j}(t)=\frac{1}{|V(t)|-1}\sum_{i=0}^{t}T(i,t).
\end{equation}

Generally speaking, the above algorithm is just available for some specific models which are built by adopting completely preferential attachment strongly dependent of vertex degree. Even so, the convenience of our algorithm should not be ignored. The $MFPT$ for random walks on a lot of famous deterministic scale-free models, including Apollonian networks \cite{Andrade-2005}, Sierpinski network \cite{Wang-2017}, are able to be facilely captured using our here thought behind the provided algorithm.

Although we have put forward algorithm 2 for $MFPT$ over the entity model $F(t)$, the closed-form expression of $MFPT$ is still not thoroughly acquired. In this case, our task is to find out other algorithms for a closed-form solution. The next subsection will focus on this problem and we gain the desired result finally.

\subsection{Algorithm 3 for the $MFPT$}

Here, we can obtain an exact solution to the $MFPT$ by induction on time step $t$. In order to achieve this task, the subsection will comprise two parts: the first is to set up a crucial recurrence equation for $T_{01}(t)$ as described in proposition 1 and the other is to analytically show a precise solution to the $MFPT$.

\textbf{Proposition 1} For any time step $t$ ($\geq 1$), the expression to the $T_{01}(t)$ satisfies

\begin{equation}\label{eqa:3-3-1}
T_{01}(t)=1+\sum_{i=1}^{t}F_{i}(1+T_{01}(t-i)).
\end{equation}

\textbf{Proof} For convenience and purpose, let us show the first several concrete forms of Eq.(\ref{eqa:3-3-1}) as $t=1, 2$.

As $t=1$, based on Eq.(\ref{eqa:3-0-2}), we may write a couple of equations

\begin{equation}\label{eqa:3-3-2}
T_{01}(1)=\frac{F_{0}}{F_{3}-1}+ \frac{F_{1}}{F_{3}-1}(1+T_{11}(1)), \quad T_{11}(1)=1+T_{01}(1).
\end{equation}

By simple computations, we obtain

\begin{equation}\label{eqa:3-3-3}
T_{01}(1)=1+F_{1}(1+1)=1+F_{1}(1+T_{01}(0))=3.
\end{equation}
here we directly use $T_{01}(0)=1$. Proposition 1 is complete.

Taking the same method as computing, $T_{01}(1)$, $T_{01}(2)$ can be expressed
\begin{equation}\label{eqa:3-3-4}
\begin{split}&T_{01}(2)=\frac{F_{0}}{F_{4}-1}+\frac{F_{1}}{F_{4}-1}(1+T_{11}(2))+\frac{F_{2}}{F_{4}-1}(1+T_{23}(2)),\\
&T_{11}(2)=\frac{F_{0}}{F_{3}-1}(1+T_{01}(2))+\frac{F_{1}}{F_{3}-1}(1+T_{25}(2)),\quad T_{25}(2)=1+T_{11}(2),\\
&T_{23}(2)=T_{24}(2)=1+T_{01}(2).
\end{split}
\end{equation}

Implementing some iterative arithmetics on Eq.(\ref{eqa:3-3-4}), we get

\begin{equation}\label{eqa:3-3-5}
T_{01}(2)=1+F_{1}(1+T_{01}(1))+F_{2}(1+T_{01}(0))=9.
\end{equation}

Obviously, the result of Eq.(\ref{eqa:3-3-5}) meets proposition 1. According to Eqs.(\ref{eqa:3-3-3}) and (\ref{eqa:3-3-5}), for any time step $l$ ($\leq t-1$), suppose proposition 1 is true. Next, for time step $t$, we will turn out proposition 1 to be correct. To do this, let us again recall the generation of model $F(t)$. For one vertex $v$ created at time step $\tau$ of the $01-root$ subtree $ST_{01}(t)$, there are $F_{t-\tau}$ vertices linked to vertex $v$. Especially, $F_{t}$ new vertices will be generated and then connected to the root vertex labeled $01$. For convenience, we here introduce some useful notations $\Lambda_{i}(t)$ $i\in[1,t]$. The $\Lambda_{i}(t)$ is regarded as the $FPT$ for a walker positioned at a vertex that is one of the neighbor set $N_{01}(t)$ of root vertex labeled $01$ and is created at time step $i$. Note that, with a such statement, the $FPT$ for a walker seated at root vertex marked $01$ obeys
\begin{equation}\label{eqa:3-3-6}
T_{01}(t)=\frac{F_{0}}{F_{t+2}-1}+\sum_{i=1}^{t}\frac{F_{i}}{F_{t+2}-1}(1+\Lambda_{i}(t)).
\end{equation}

Due to self-similarity of Fibonacci tree $F(t)$, we may have

\begin{equation}\label{eqa:3-3-7}
\Lambda_{i}(t)=T_{01}(t)+T_{01}(t-i).
\end{equation}

To better understand Eq.(\ref{eqa:3-3-7}), we will explain this in detail as follows. If we think of the root vertex $01$ as a supposed destination point of model $F(t)$, then one subtree generated by some vertex $u$, which is added at time step $i$ into its neighbor set $N_{01}(t)$, can possess similar outline to $01-root$ subtree $ST_{01}(t-i)$. Accordingly, the $FPT$ for a walker allocated at vertex $u$ to visit the root vertex $01$ follows $T_{01}(t-i)$ as the second term in the right hand side of Eq.(\ref{eqa:3-3-7}). There is a unique path connecting vertex $u$ to destination point $00$, since vertex $u$ must first arrive at vertex $01$ at cost of $T_{01}(t-i)$ and final hops on vertex $00$ taking $T_{01}(t)$ as the first term in the right hand side of Eq.(\ref{eqa:3-3-7}). This completes the proof of Eq.(\ref{eqa:3-3-7}).

Inserting Eq.(\ref{eqa:3-3-7}) into Eq.(\ref{eqa:3-3-6}) has the desired result as depicted in Eq.(\ref{eqa:3-3-1}). Therefore, we validate proposition 1 by both induction and hypothesis.

\begin{figure}
\centering
  \includegraphics[height=5cm]{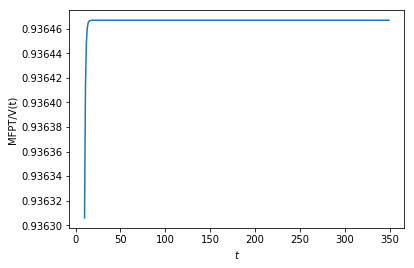}\\
{\small Fig.5. The diagram of the $MFPT/|V(t)|$ on Fibonacci tree $F(t)$. Although here shows that the $MFPT$ is linearly correlated with the order $|V(t)|$ as $t\geq20$. The solution to $MFPT$ in fact contains another component, $|V(t)|^{\frac{\ln2}{\ln(1+\sqrt{3})}-1}$, which be dominated by $|V(t)|$ because of $\frac{\ln2}{\ln(1+\sqrt{3})}-1< 1$ as $t\rightarrow\infty$.  }
\end{figure}

Based on the initial conditional value $T_{01}(0)=1$ and many methods for non-homogeneous recurrence equations, the closed form of $T_{01}(t)$ can be written

\begin{equation}\label{eqa:3-3-8}
T_{01}(t)=\frac{2+\sqrt{3}}{3}(1+\sqrt{3})^{t}+\frac{2-\sqrt{3}}{3}(1-\sqrt{3})^{t}-\frac{1}{3}.
\end{equation}

After that, the left task is to analytically calculate an exact expression to the $MFPT$. Compared with the description of Eq.(\ref{eqa:3-2-7}), we intend to introduce another algorithm highly correlated to the underlying structure of Fibonacci tree $F(t)$. To organize the following portion smoothly, let us take a group of notations $\Gamma(i)$ ($i\in [0,t]$). Our model $F(t)$ can be restructured by number $F_{0}$ $01-root$ subtree $ST_{01}(t)$, $F_{1}$ $01-root$ subtrees $ST_{01}(t-1)$, $F_{2}$ $01-root$ subtrees $ST_{01}(t-2)$s, ... , $F_{t}$ $01-root$ subtrees $ST_{01}(0)$ in a manner that connects vertex $00$ to root vertex of each subtree $ST_{01}(i)$ ($i\in [0,t]$) mentioned above using a new edge. The total number of first-passage time which each vertex of subtree $ST_{01}(i)$ ($i\in [0,t]$) first arrive at vertex $00$ can be viewed as $\Gamma(i)$, hence Eq.(\ref{eqa:3-0-3}) can be reorganized as

\begin{equation}\label{eqa:3-3-9}
MFPT_{t}=\frac{1}{|V(t)|-1}\sum_{j\in V(t), j\neq o}L_{j}(t)=\frac{1}{|V(t)|-1}\sum_{i=0}^{t}F_{i}\Gamma(t-i).
\end{equation}

Computing $MFPT_{t}$ over the entire model $F(t)$ may be naturally reduced to look for a solution to $\Gamma(i)$. The key point in such a process is to build up a relationship among individual $\Gamma(i)$. In view of the declaration proving the validity of Eq.(\ref{eqa:3-3-7}), one may capture

\begin{equation}\label{eqa:3-3-10}
\Gamma(t)=\frac{|V(t)|}{2}T_{01}(t)+\sum_{i=1}^{t}F_{i}\Gamma(t-i).
\end{equation}

Implementing some simple calculations to resolve Eq.(\ref{eqa:3-3-10}) produces

\begin{equation}\label{eqa:3-3-11}
\begin{split}\Gamma(t)&=\frac{b_{t-1}}{2}\Gamma(1)+b_{t}\Gamma(0)+\frac{|V(t)|}{2}T_{01}(t)-\frac{|V(t-1)|}{2}T_{01}(t-1)-\frac{|V(t-2)|}{2}T_{01}(t-2)\\
&\quad+\sum_{i=2}^{t-1} b_{t+2-i}\left[\frac{|V(i)|}{2}T_{01}(i)-\frac{|V(i-1)|}{2}T_{01}(i-1)-\frac{|V(i-2)|}{2}T_{01}(i-2)\right]
\end{split}
\end{equation}
here $\Gamma(0)=1$, $\Gamma(1)=7$ and $b_{t}=\frac{3+\sqrt{3}}{3}(1+\sqrt{3})^{t-2}+\frac{3-\sqrt{3}}{3}(1-\sqrt{3})^{t-2}$.

Substituting Eq.(\ref{eqa:3-3-11}) into Eq.(\ref{eqa:3-3-9}), we have the next theorem

\textbf{Theorem 2} The solution of the mean first-passage time ($MFPT$) for trapping point located at either of two hubs of Fibonacci tree $F(t)$ is
\begin{equation}\label{eqa:3-3-12}
MFPT_{t}=\frac{1}{|V(t)|-1}\sum_{i=0}^{t}F_{i}\Gamma(t-i)\propto |V(t)|+|V(t)|^{\frac{\ln2}{\ln(1+\sqrt{3})}-1} .
\end{equation}
From the provided results above, we attain a rigorous solution to the $MFPT_{t}$ over the entire model $F(t)$.

\textbf{NOTE } Surprisingly, the $MFPT_{t}$ is not linearly correlated with the order of Fibonacci tree $F(t)$. In some extent, our result is not consistent enough with an assertion, namely $MFPT\sim |V|$, such as, recurrence trees \cite{Bollt-2005}. Hence, our Fibonacci trees $F(t)$ indeed enriches research in this field with its own advantage. Perhaps such a finding brings significant insight into better studying diffusion in complex networks and then makes people's understanding about random walks on treelike models more comprehensive. As shown in Fig.5, for the large value of $t$, the first term performs an absolute superiority and ultimately dominates the convergence rate of the $MFPT_{t}$. To highlight contribution to the $MFPT_{t}$ from the second term in Eq.(\ref{eqa:3-3-12}), we apply an approximate estimate for the coefficient of $|V(t)|$ from Fig.5 to define

\begin{equation}\label{eqa:3-3-13}
\theta(t): \frac{\ln (MFPT_{t}-\phi|V(t)|)}{\ln|V(t)|}.
\end{equation}

It is noteworthy that the outline of Fig.6 has considerable agreement with Eqs.(\ref{eqa:3-3-12}) and (\ref{eqa:3-3-13}). Therefore, our statements are strongly validated by the performance of numerical experiments.

As known, the $MFPT$ is a useful topological structure parameter portraying the efficiency of information diffusion over network. Clearly, Eq.(\ref{eqa:3-3-12}) reports that in the large graph size limit, the $MFPT$ not only increases with the vertex number $|V(t)|$ of model $F(t)$ but also grows as a power-law function of $|V(t)|$ with exponent larger than $1$ and less than $2$. Compared to the $OMFPT$ discussed in subsection 3.1, the order of models $F(t)$ plays a more important role than its diameter. We here think that a plausible reason for this is how much information a walker does really know about network structure. If he has known a shortest path connecting his current position to his destination point (trapping point), then he takes useful advantage of the information to arrive at that trapping point rapidly. Strictly speaking, the diameter is no less than arbitrary shortest path, indirectly suggesting which the diameter has considerable influence on the $OMFPT$. On the contrary, before starting with walking on a network model, there are no knowledge about the underlying structure of network disclosed to a walker and then he has to select with equal probability one vertex from his neighbor set as his next position. The phenomena usually leads to a result that it is a higher possibility for a walker to traverse almost all vertices on models $F(t)$ before ultimately reaching his destination point. Such an explanation can be validated by many previous proposed models including Apollonian networks \cite{Andrade-2005}, Sierpinski network \cite{Wang-2017}, Recursive tree network \cite{Kursten-2016}, and T-fractal \cite{Agliari-2008}. Another fact is that the expression in Eq.(\ref{eqa:3-3-12}) is different from those previously proposed viewpoints that nearly say that the $MFPT$ for random walks on a scale-free tree should better agree with a linear relationship to its vertex number. Indeed, this difference between them is particularly because Fibonacci trees $F(t)$ can be built in a distinct growth manner, namely, \emph{non-monotone preferential attachment}. This is exactly one of our contributions in this paper.

\begin{figure}
\centering
  \includegraphics[height=4cm]{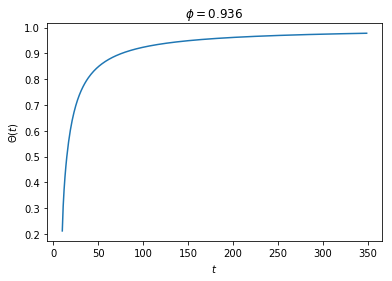}\quad \includegraphics[height=4cm]{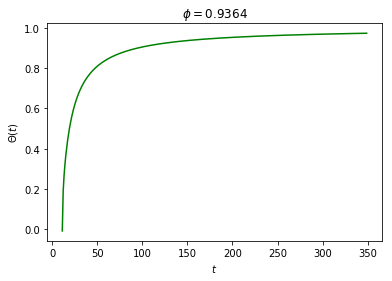}\quad \includegraphics[height=4cm]{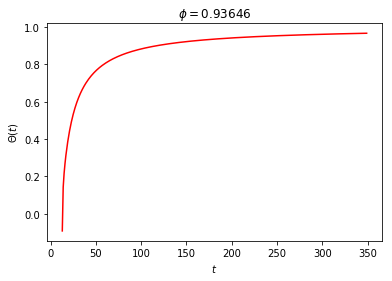}\\
{\small Fig.6. The diagram of $\theta(t)$. It appears to show that each solid line is not perfectly consistent with our analytical result, i.e., the existence of the power-law term $|V(t)|^{\frac{\ln2}{\ln(1+\sqrt{3})}-1}$ in the $MFPT$. One of the most prominent reasons still is the first term of Eq.(\ref{eqa:3-3-12}) plays a vital role on the expression of the $MFPT$.     }
\end{figure}

\section{Conclusion}

In summary, we in this paper propose a family of new network models, named Fibonacci trees $F(t)$, and study in-depth two kinds of representations corresponding to discrete Markov process. In view of the intrinsic property of Fibonacci sequence, our models $F(t)$ are proved to display the scale-free feature. Differed with pre-existing growth manners for generating deterministic complex network models, we require that the number of newly added vertex connected to an older vertex be highly consistent with degree change of the older vertex not with its current degree valve. Subsequently, we discuss the optimal mean first-passage time ($OMFPT$) for biased Markov random walk on models $F(t)$ and obtain a precise solution to the $OMFPT$ using algorithm 1 based on self-similarity prevailing in most complex networks \cite{Song-2005}. In order to study the mean first-passage time ($MFPT$) for unbiased Markov random walk on models $F(t)$, we introduce both algorithms. The first is to output an approximate solution and may be suitable to a lot of models as mentioned in Section 3. The other is for calculating a deterministic solution by taking into account model itself underlying structure. We hope that the thoughts reflected by the second algorithm can open a new approach to investigate the unbiased Markov random walk on deterministic models that possess similar properties to Fibonacci trees $F(t)$. One of reasons for this is to utilize a trivial yet useful method, i.e., recurrence equations, to calculate a closed-form solution to the $MFPT$.

\emph{Problem 1}. As reported in our work, it is supposed that both Markov random walks on models $F(t)$ arise in extremely ideal environment. The former shows an assumption where a walker completely learns about the entire network structure and always chooses the optimal path to arrive at destination point. Although this is very efficient for a walker, the probability of causing congestion becomes quite high if a amount of walkers under a network all do so, resulting in which the whole network will soon be in an avalanche. On the other hand, the unbiased Markov random walk hints a fact that there is no information about topological structure of networks in question sent to a walker. This commonly lows down the transmit efficiency of information diffusion. Taking into account small-world property prominent in real-world complex networks, namely, $D\sim \ln(|V|)$ or $D\sim \ln\ln(|V|)$, The mean first-passage time for random walks may be transformed from a phase, linear correlation with the diameter $D$, to another pole, power-law correlation with the order $|V|$, when changing probability for selecting the next candidate node from the optimal to the uniform. In reality, the truth transmit process is nor the former neither the later. A walker at least knows about parts of knowledge over network structure. To better stimulate this scenario, the supposed probability should meet some type of distribution form. Such an assumption is that a walker positioned at vertex $u$ is able to choose his next hopping point $v$ using the probability $P_{u\rightarrow v}=\frac{k_{v}}{\sum_{i\in N_{u}}k_{i}}$ similar to that for constructing the BA-model. We believe that the above discussions will be of great interest in the future. Noticeably, percolation and phase transform must be two attractive topics in such an assumption probability. Therefore, we will continuously focus on research in this field and refer to the above research as our future task.

\begin{figure}
\centering
  \includegraphics[height=5cm]{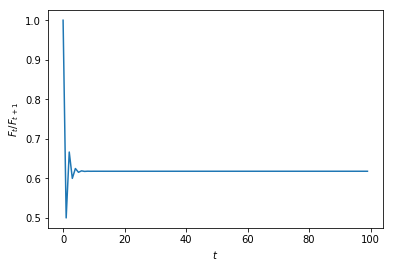}\\
{\small Fig.7. The diagram of the ratio $F_{t}/F_{t+1}$ of Fibonacci sequence $\{F_{t}\}$.  }
\end{figure}

\emph{Problem 2}. As addressed previously, almost all complex network models have been constructed using the \emph{monotone preferential attachment}. Based on network models of this type, many interesting topics have been discussed in detail. Examples of such interests include synchronization phenomena, random walks, percolation and system control (attack and resistance). Among them, the power-law exponent $\gamma$ of deterministic network models is usually less than $3$ and the $MFPT$ for random walks on scale-free tree model does linearly correlate to its order. By contrast, we here propose Fibonacci trees $F(t)$ without the need to comply with the \emph{monotone preferential attachment}. Fibonacci trees $F(t)$ are generated with respect to Fibonacci sequence $\{F_{t}\}$. Since model $F(t)$ inherits an important character from Fibonacci sequence $F_{t}$ in which the ratio $F_{t}/F_{t+1}$ ($t\geq0$) has a non-monotone tendency to a constant $0.618$, see Fig.7. Due to the trivial but necessary property, model $F(t)$ reports some intriguing results. Model $F(t)$ not only is scale-free as predicted but also has a power-law parameter $\gamma$ larger than $3$. Meanwhile, as said in Eq.(\ref{eqa:3-3-12}), to some extent, the $MFPT$ for random walks on model $F(t)$ does not increase linearly with its vertex number rather than is a combination form covering both a linear component and a power-law one. The both unexpected achievements are precise in terms of the \emph{non-monotone preferential attachment} used to create model $F(t)$. From the both theoretical and practical point of view, we hope our works can help people to deeply understand the fundamental mechanism over dynamical complex networks. We sincerely believe that our work is just a tip of the iceberg. In order to better understand complex networks, there will be a lot of challenging and glamorous works to do in the future.

\vskip 1cm

\textbf{Acknowledgment.} We thank Wang Xiaomin for some useful discussions. The research was supported by the National Key Research and Development Plan under grant 2017YFB1200704, and the National Natural Science Foundation of China under grants No. 61662066 and  No. 61672059.

\vskip 1cm

\textbf{Complemental Material}

\textbf{Appendix A} Definition of the Fibonacci sequence $\{F_{t}\}$

A sequence consisting of positive integers $F_{i}$ ($i=0, 1, 2, ...$) can be called Fibonacci series, if the first three terms obey $F_{0}=1$, $F_{1}=1$, $F_{2}=2=F_{0}+F_{1}$ and the forward elements ($i\geq 3$) all satisfy

\begin{equation}\label{eqa:mf-2-1}
F_{i}=F_{i-1}+F_{i-2}.
\end{equation}

Sometime this sequence is also vividly referred to as ¡°\emph{Rabbit Sequence}¡±.

\textbf{Appendix B} Description of the $t$-dimensional square matrix $\mathbf{F^{\Delta}(t)}$

$$\mathbf{F^{\Delta}(t)}=\left(
  \begin{array}{ccccccc}
    F_{1} & F_{2} & F_{3} & ...& F_{t-2} & F_{t-1} & F_{t} \\
    0 & F_{1} & F_{2} & ... & F_{t-3}& F_{t-2} & F_{t-1} \\
     0 & 0 & F_{1} &... & F_{t-4}& F_{t-3} & F_{t-2} \\
    \vdots & \vdots & \vdots & \ddots & \vdots & \vdots & \vdots  \\
     0 & 0 & 0 & ... & F_{1} & F_{2} & F_{3}\\
    0 & 0 & 0 & ... & 0& F_{1} & F_{2} \\
    0 & 0 & 0 & ... & 0 & 0 & F_{1} \\
  \end{array}
\right)_{t\times t}$$

\end{document}